\documentclass[%
 reprint,
showpacs,preprintnumbers,
 amsmath,amssymb,
]{revtex4-1}

\usepackage{graphicx}% Include figure files
\usepackage{multirow}

\newcommand{\myfig}[4][ht]{
\begin{figure}[#1]
\centering
\includegraphics[#2]{#3}
\caption{#4\label{#3}}
\end{figure}
}

\newcommand{\myfigwide}[4][ht]{
\begin{figure*}[#1]
\centering
\includegraphics[#2]{#3}
\caption{#4\label{#3}}
\end{figure*}
}

\graphicspath{{images/}}

\usepackage{dcolumn}% Align table columns on decimal point
\usepackage{bm}% bold math

\begin{document}

%\preprint{Preprint submitted to XXX}

\title{Limitations of generalised grey phonon models for\\quasiballistic thermal transport in time-periodic regimes}
\author{Bjorn Vermeersch$^1$}
\email{bjorn.vermeersch@cea.fr}
\author{Ali Shakouri$^2$}
\affiliation{\vspace{3mm}$^1$ CEA, LITEN, 17 Rue des Martyrs, 38054 Grenoble, France\\$^2$ Birck Nanotechnology Center, 1205 West State St, West Lafayette IN 47907, Indiana USA}
\date{\today}
\begin{abstract}
Suitably superimposed grey-medium solutions of the Boltzmann transport equation (BTE) provide a simple yet accurate description of non-grey quasiballistic heat conduction in transient thermal grating experiments. Recent applications of similar strategies based on kinetic and McKelvey-Schockley-Landauer theory to time-periodic transport predicted notable conductivity suppression only at heating frequencies comparable to phonon scattering rates, in contrast to lengthscale criteria observed by several prior studies. Here we show that the frequency-integrated grey-medium approximation (FIGMA) is ill suited to tackle temporally periodic quasiballistic transport. Starting from first-principles phonon dispersions and scattering rates, we obtain semi-analytic 1D BTE solutions for semi-infinite structures subjected to sinusoidal surface heating and compare these to the approximate model counterparts. We find FIGMA-based approaches to overestimate the semiconductor surface temperature by up to one and characteristic heating frequencies for onset of quasiballistic effects by up to three orders of magnitude respectively. Our study reasserts that experimentally observed heating-frequency dependent apparent conductivities originate in the overlap of the characteristic length scale of the thermal gradient with phonon mean free paths.
\end{abstract}
\maketitle
%\tableofcontents
\section{Introduction}
Experimental evidence of quasiballistic phonon transport in temporally periodic heating regimes was first reported in 2007 by Koh and Cahill \cite{cahill}. The thermal conductivity of semiconductor alloys measured by time-domain thermoreflectance (TDTR) decreased with laser modulation frequency by nearly 50\% over the 1--10$\,$MHz range. Regner and coworkers \cite{malen} observed similar behaviour in frequency domain thermoreflectance (FDTR) on Si, though this result was later suggested to be an interpretation artifact caused by the complicated heat flow in the Au/Cr transducer used in FDTR \cite{wilson}.
\par
The effect in alloys was originally attributed to the notion that phonons with mean free paths (MFPs) exceeding the characteristic length scale of the induced thermal gradient do not contribute to the experimentally observed thermal conductivity \cite{cahill}. Alternative explanations were explored by several subsequent works. Wilson and coworkers \cite{twochannel} analysed the problem in terms of nonequilibrium transport within a two-channel configuration. Koh and coworkers \cite{kohnonlocal} connected the observed behaviour to the nonlocality of the constitutive law between heat flux and temperature gradient. Vermeersch and coworkers \cite{levy1,levy2} explained the effect from first principles through the presence of fractal L\'evy transport dynamics. Interestingly, a common aspect that unifies these diverse perspectives is that each deviates from regular diffusive theory through a fundamental alteration of the spatial signature of the thermal fields. Within these viewpoints, notable conductivity suppression is expected for phonons whose MFP $\Lambda$ approach or exceed the source thermal penetration length $\ell(\omega_{\text{H}}) = \sqrt{2D/\omega_{\text{H}}}$ with $\omega_{\text{H}}$ the angular heating frequency and $D = \kappa/C$ the bulk diffusivity of the medium.
\par
Recently, Yang \& Dames \cite{ref1} and Maassen \& Lundstrom \cite{ref2} independently theorised that quasiballistic deviations instead occur when the heating frequency becomes comparable to phonon scattering rates $\tau^{-1}$. Both of these works operated under the so called frequency-integrated grey medium approximation (FIGMA). This framework has been successfully applied to transient thermal grating (TTG) problems \cite{collins,zeng} and consists of determining approximate thermal dynamics of realistic (non-grey) crystals by superimposing exact grey (single MFP) solutions of the Boltzmann transport equation (BTE). Yang and Dames directly focused on the apparent thermal conductivity and used kinetic theory reasoning to extend their grey-medium result $\kappa_{\text{app}}^{\text{gr}}(\omega_{\text{H}})$ to non-grey crystals \cite{ref1}. Maassen and Lundstrom, meanwhile, employed a Landauer-type perspective to explore temperature and heat flux fields in temporally periodic regime by integrating previously obtained McKelvey-Shockley grey solutions \cite{maassen_transient} over phonon energy \cite{ref2}.
\par
The onset for notable heating-frequency dependent effects $\omega_{\text{H}} \tau \geq 1$ derived under the FIGMA, which is reminiscent of a similar timescale criterion determined by Volz \cite{volz}, poses an intriguing and somewhat puzzling contrast with the aforementioned lengthscale criterion suggested by several prior studies. Here, we investigate the situation by analysing semi-infinite semiconductors with first-principles phonon properties under periodic surface heating. Comparing the approximate approaches (summarised in Section \ref{sec:FIGMAreview}) directly to non-grey BTE solutions (outlined in Section \ref{sec:BTE}) reveals that FIGMA models are poorly suited to describe temporally periodic quasiballistic transport (Section \ref{sec:results}), despite their previously validated performance in other settings (Section \ref{sec:discussion}). A short summary (Section \ref{sec:conclusions}) concludes the paper.
\section{Brief review of FIGMA solutions}\label{sec:FIGMAreview}
\subsection{Apparent conductivity (kinetic theory)}
Yang and Dames \cite{ref1} derived the exact grey BTE solution for a semi-infinite geometry with a two-flux approach. Observing that the heat flux and temperature gradient possess identical spatial signatures, they obtained the apparent conductivity as
\begin{equation}
\kappa_{\text{app}}^{\text{gr}}(\omega_{\text{H}}) = \left\| \frac{q^{\text{gr}}(x;\omega_{\text{H}})}{- \frac{\partial \Delta T^{\text{gr}}(x;\omega_{\text{H}})}{\partial x}} \right\| = B^{\text{gr}}(\omega_{\text{H}} \tau) \, \kappa \label{kappagreyYD}
\end{equation}
The function $B$, which captures the suppression of the nominal Fourier conductivity $\kappa$ due to heating-frequency dependent quasiballistic effects, is a relatively complicated expression of 4 variables that each depend on $\omega_{\text{H}} \tau$ themselves \cite{ref1}. However, upon closer inspection we found that the published solution actually reduces \textit{exactly} to
\begin{equation}
\kappa_{\text{app}}^{\text{gr}}(\omega_{\text{H}}) = \frac{\kappa}{\sqrt{1+\omega_{\text{H}}^2 \tau^2}} \label{kappaeff_grey}
\end{equation}
This result can in fact be obtained far more directly by solving the semi-infinite BTE in transformed domains (Appendix \ref{app:greymedia}). The grey solution is then extended approximately to multimodal media through kinetic theory arguments \cite{ref1}. Formulated for a crystal supporting an array of discrete phonon channels with heat capacities $C_n$, group velocities $\vec{v}_n$ and MFPs $\Lambda_n$, we have
\begin{equation}
\kappa_{\text{app}}(\omega_{\text{H}}) = \sum \kappa_{\text{app},n}^{\text{gr}}(\omega_{\text{H}}) = \sum \frac{C_n \, v_n \, \Lambda_n \, \cos^2 \theta_n}{\sqrt{1+\omega_{\text{H}}^2 \tau_n^2}} \label{kappaeffFIGMA}
\end{equation}
where $\theta$ denotes the angle the group velocity makes with the 1D transport axis. Notice this model captures `strongly quasiballistic' (short time scale) effects but lacks the `weakly quasiballistic' regime (long time scales but length scales comparable to MFPs) inherently present in the nongrey BTE \cite{minnich}.
\subsection{Thermal fields (Landauer approach)}
Maassen and Lundstrom \cite{ref2} demonstrated that heat conduction in a grey medium obeys the hyperbolic heat equation at all length and time scales and then extend the solutions to non-grey crystals by phonon frequency integration. Formulated for a discrete set of channels, the total temperature field is approximated as  
\begin{equation}
\Delta T = \frac{P}{C} = \frac{\sum \frac{C_n}{C} \, P_n^{\text{gr}}}{\sum C_n} \label{deltaTFIGMA}
\end{equation}
where $P$ signifies deviational thermal energy per volume unit. Notice that within this viewpoint each channel independently searches equilibrium with its own pseudo-temperature $\Delta T_n^{\text{gr}} = P_n^{\text{gr}}/C_n$. The non-grey BTE, by contrast, is governed by search for equilibrium with one universal temperature $\Delta T$ (no subscript) as expressed by the energy conservation equation $\sum (P_n - C_n \, \Delta T)/\tau_n = 0$ \cite{levy1,minnich}. This leads to
$\Delta T = [\sum P_n / \tau_n]/[\sum C_n / \tau_n]$, which clearly differs from (\ref{deltaTFIGMA}) through the presence of $\tau_n$ and the fact that generally speaking $P_n \neq (C_n/C) \, P_n^{\text{gr}}$.
\par
Using the grey solution from Ref. \onlinecite{ref2}, the energy density inside a semi-infinite medium induced by a 1$\,$W/m$^2$ surface heat flux at angular frequency $\omega_{\text{H}}$ obeys
\begin{eqnarray}
P_{\text{Landauer}}(x;\omega_{\text{H}}) & = &\sum \limits_{\text{$\omega$ bins}} \frac{C_{\omega}}{C} \cdot \frac{\gamma_{\omega} \exp \left( - \beta_{\omega} \, x \right)}{\kappa_{\omega} \, \beta_{\omega}} \label{deltaT_FIGMA} \\
\text{with} \quad \gamma_{\omega} & = & 1 + j \, \frac{4}{3} \, \omega_{\text{H}} \, \tau_{\omega} \quad , \quad \beta_{\omega} = \sqrt{\frac{j\omega_{\text{H}} \, \gamma_{\omega}}{D_{\omega}}} \nonumber
\end{eqnarray}
and $\kappa_{\omega} = C_{\omega} \, D_{\omega} = C_{\omega} \, v_{\omega} \, \Lambda_{\omega}/3$ the spectral conductivity. The factor $4/3$ in $\gamma_{\omega}$ arises from mapping 3D isotropic phonon motion to 1D transport, as explained in Ref. \onlinecite{ref2}.
\section{BTE modeling}\label{sec:BTE}
We will concentrate on semi-infinite structures with temporally periodic heat source at the top surface. While we are not aware of any prior explicit BTE analyses of this particular configuration, all essential elements for deriving and validating semi-analytic non-grey solutions are available in the current literature as outlined below. We carry out all of our calculations under the relaxation time approximation (RTA) with first-principles phonon dispersions and scattering rates.
\subsection{First-principles phonon properties}
We start by computing \textit{ab-initio} interatomic force constants and associated phonon properties devoid of any adjustable parameters for Si, Si$_{0.4}$Ge$_{0.6}$, Si$_{0.82}$Ge$_{0.18}$ and In$_{0.53}$Ga$_{0.47}$As through a well established framework documented elsewhere \cite{shengBTE,abinitiobookchapter,virtualcrystal1,virtualcrystal2}. We perform our DFT calculations exactly as described in Ref. \onlinecite{levy1}. Briefly, we carry out unconstrained unit cell relaxations with \textsc{VASP} \cite{vasp} under the LDA \cite{lda} with energy cutoff 30\% above the pseudopotential maximum. We use $5 \times 5 \times 5$ supercells for computing the second- and third-order force constants, the latter of which include the effects of the 5 nearest neighbours. Coulomb interactions in polar compounds are accounted for through effective Born charges \cite{born}.
\par
The key outcome is a set $\{ C_k,\vec{v}_k,\tau_k \}$ of heat capacities, group velocities and relaxation times resolved over a 3D discretisation of the Brillouin zone. Here the generalised index $k$ labels for both wavevector and phonon branch. Spectrally resolved parameters, used to evaluate the Landauer solutions (\ref{deltaT_FIGMA}), are readily obtained through phonon frequency binning. Resulting bulk thermal properties, dispersions and cumulative conductivity curves are provided in Appendix \ref{app:abinitiodata} for benchmarking convenience.
\par
The computed first-principles phonon properties offer convenient and fairly realistic inputs to the various thermal models being investigated here. It must be noted that minor inaccuracies which inevitably remain within the first-principles data do not affect the central outcomes of this work in any way. The key observation to be emphasized here is that, given the same set of phonon inputs, FIGMA solutions display severe qualitative and quantitative discrepancies from the BTE counterparts they purportedly approximate.
\subsection{Thermal fields}
Analytic solutions for the 1D BTE in fully infinite isotropic media were derived by Hua and Minnich \cite{minnich} and then generalised to crystals with arbitrary anisotropy by Vermeersch and coworkers \cite{levy1}. Extension to semi-infinite geometries is non-trivial because phonons that hit the top surface can scatter randomly into a multitude of modes that all obey the boundary condition (details in Appendix \ref{app:VRMC}). However, we have verified that variance-reduced Monte Carlo simulations of semi-infinite structures produce transient temperature fields that are virtually indistinguishable from infinite-medium solutions upscaled by a factor of two (see Appendix \ref{app:VRMC} as well). With minimal loss of accuracy, we can therefore perform all intermediate BTE calculations assuming infinite media, and then simply double the thermal fields at the end. For FIGMA models the scaling factor of two is exact, since the thermal field in a semi-infinite grey medium is \textit{precisely} twice that of an infinite one at all length and time scales (see Appendix \ref{app:greymedia}).
\par
The weakly quasiballistic single pulse response of the RTA-BTE takes the functional form \cite{levy1} 
\begin{equation}
P(\xi,s) = \frac{1}{s + \psi(\xi)} \quad \leftrightarrow \quad P(\xi,t) = \exp \left[ -\psi(\xi) \, t \right] \label{PBTE1}
\end{equation}
where $\xi$ denotes spatial frequency and $s$ is the Laplace variable. The propagator function $\psi(\xi)$ is directly connected to the first-principles phonon properties as
\begin{equation}
\psi(\xi) = \sum \frac{C_k \, \xi^2 \Lambda_{x,k}^2}{\tau_k [1 + \xi^2 \Lambda_{x,k}^2]} \,\, \biggr / \sum \frac{C_k}{1 + \xi^2 \Lambda_{x,k}^2} \label{PBTE2}
\end{equation}
While we used the exact expression for $\psi(\xi)$ in all computations, it is worth noting that thermal transport in an alloy compound can be accurately described by its nominal Fourier diffusivity $D$, L\'evy exponent $\alpha$ (usually $\simeq 1.7$) and diffusive recovery length $x_{\text{R}}$ (typically a few microns) through the compact form $\psi(\xi) \simeq D\xi^2/(1+x_{\text{R}}^2 \xi^2)^{1-\alpha/2}$.
\par
The solution (\ref{PBTE1}) ignores purely ballistic transport effects but offers excellent performance at temporal scales exceeding characteristic phonon relaxation times, which are typically below 1$\,$ns (see Appendix \ref{app:abinitiodata}). Our BTE solutions for periodic regimes therefore apply across the entire experimentally achievable bandwidth $f_{\text{H}} \lesssim 200\,$MHz.
\par
Fourier inversion to real space
\begin{equation}
P(x,t|s) = \frac{1}{\pi} \, \int \limits_{0}^{\infty} P(\xi,t|s)\,\cos(\xi x) \, \mathrm{d}\xi \label{PBTE3}
\end{equation}
can be performed semi-analytically by using that
\begin{multline}
x \neq 0 : \int (A_0 + A_1 \xi + A_2 \xi^2) \, \cos(\xi x) \, \mathrm{d}\xi = - \frac{2 A_2 \sin(\xi x)}{x^3} \\ 
+ \frac{(A_1 + A_2 \xi) \, \cos(\xi x)}{x^2} + \frac{(A_0+ A_1 \xi + A_2 \xi^2) \sin(\xi x)}{x} \label{primitiveinvfourier}
\end{multline}
For time domain responses, used for comparison with Monte Carlo simulations in Fig. \ref{fig6-VRMC}, we perform a piecewise Taylor series expansion over consecutive $\xi$ intervals
\begin{equation}
\exp[-\psi(\xi) \, t] \simeq A_{0,n}(t) + A_{1,n}(t) \, \xi + A_{2,n}(t) \, \xi^2
\end{equation}
In periodic regime, on the other hand, we have
\begin{equation}
P(\xi,s=j\omega_{\text{H}}) = \frac{1}{j \omega_{\text{H}} + \psi(\xi)} = \frac{\psi(\xi) - j \omega_{\text{H}}}{\psi^2(\xi) + \omega_{\text{H}}^2} 
\end{equation}
%Given that
%\begin{equation}
%\frac{\mathrm{d}P_r}{\mathrm{d}\xi} = \frac{P_r}{\psi} \cdot \frac{\mathrm{d}\psi}{\mathrm{d}\xi} \cdot (1 - 2 \, \psi \, P_r) \,\, ; \,\, \frac{\mathrm{d}P_i}{\mathrm{d}\xi} = \frac{2 \psi}{\omega_{\text{H}}} \cdot \frac{\mathrm{d}\psi}{\mathrm{d}\xi} \cdot P_i^2
%\end{equation}
Piecewise linear Taylor series expansion again enables analytic integration via (\ref{primitiveinvfourier}). We employed a logarithmically spaced $\xi$ grid ranging from $10^{-3}\,$m$^{-1}$ to $10^{10}\,$m$^{-1}$ with $10^4$ points in our calculations. We verified that for purely diffusive transport $\psi(\xi) = D\xi^2$ our computation scheme reproduces the exact Fourier solution
\begin{equation}
P_{\text{Fourier}}(x;\omega_{\text{H}}) = \frac{\exp \left[ - \sqrt{2j} \, x/\ell \right]}{\sqrt{4j \omega_{\text{H}} D}} \quad , \quad \ell = \sqrt{\frac{2D}{\omega_{\text{H}}}} \label{PFourier}
\end{equation}
within 0.05\% in magnitude and 0.04 degrees of phase for heating frequencies $f_{\text{H}} = \omega_{\text{H}}/2\pi$ up to 1$\,$GHz.
\subsection{Apparent conductivity}
Experiments have typically no access to the internal thermal fields just discussed, but only probe the semi-infinite medium's surface response
\begin{equation}
P_0(\omega_{\text{H}}) = \frac{2}{\pi} \, \int \limits_{0}^{\infty} \frac{\mathrm{d}\xi}{j\omega_{\text{H}} + \psi(\xi)}
\end{equation}
Observing that $\int \mathrm{d}\xi/(j\omega_{\text{H}}+a+b\,\xi) = \ln(j\omega_{\text{H}} + a + b\,\xi)/b$ we can again integrate semi-analytically to find
\begin{equation}
P_0(\omega_{\text{H}}) \simeq \frac{2}{\pi} \sum \limits_{n} \ln \left[ \frac{j \omega_{\text{H}}+\psi_{n+1}}{j \omega_{\text{H}}+\psi_{n}}\right] \cdot \frac{\xi_{n+1} - \xi_n}{\psi(\xi_{n+1})-\psi(\xi_{n})} \label{P0scheme}
\end{equation}
Applying this scheme to $\psi = D \xi^2$ with the same logarithmic $\xi$ grid as above reproduces the exact diffusive solution $P_0(\omega_{\text{H}}) = 1/\sqrt{j\omega_{\text{H}}D}$ within 0.1\% in magnitude and 0.06 degrees in phase. The Fourier expression furthermore enables us to evaluate the heating-frequency dependent apparent diffusivity as%
\myfigwide[!htb]{width=0.94\textwidth}{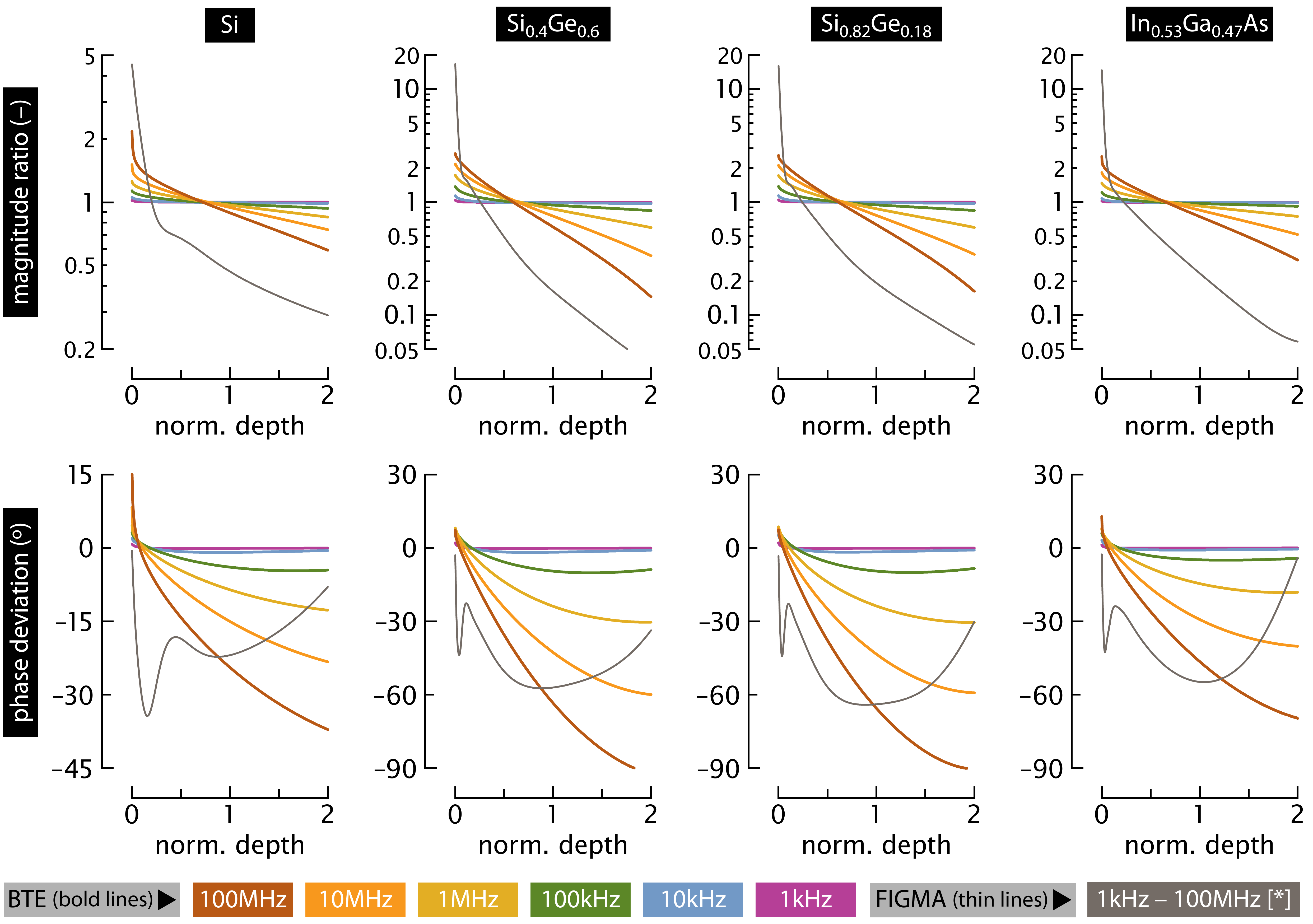}{Thermal fields in semi-infinite media with periodic surface heating. Results are plotted relative to the exact diffusive solutions versus normalised depth $x/\ell(\omega_{\text{H}})$. [*] FIGMA curves for different heating frequencies are not truly identical but differ by amounts so minute ($\leq$1.3\% in magnitude and $\leq$0.5 degrees of phase) that they are indistinguishable in the graph.}%
\begin{equation}%
\text{BTE:} \quad D_{\text{app}}(\omega_{\text{H}}) \equiv \frac{\kappa_{\text{app}}(\omega_{\text{H}})}{C} = \left \| \frac{1}{j\omega_{\text{H}} \, P_0^2(\omega_{\text{H}})} \right \| \label{kappaeffBTE}
\end{equation}%
One might argue that this prevents direct comparison with the approximate kinetic theory result (\ref{kappaeffFIGMA}) since for the latter the apparent conductivity of an individual phonon channel was determined from the relation between heat flux and temperature gradient instead of the surface response. However, both definitions are formally equivalent in grey media, as we have (see Appendix \ref{app:greymedia}):
\begin{equation}
P_0^{\text{gr}}(s) = \frac{2}{\pi} \int \limits_{0}^{\infty} \frac{(1+s\tau) \, \mathrm{d}\xi}{s\, (1+s\tau) + D \xi^2} = \frac{\sqrt{1+s\tau}}{\sqrt{sD}}
\end{equation}
In periodic regime $s = j \omega_{\text{H}}$, (\ref{kappaeffBTE}) produces $D_{\text{app}}(\omega_{\text{H}}) = D / \| 1+j\omega_{\text{H}}\tau\|$, in exact agreement with (\ref{kappaeff_grey}).
\section{Results}\label{sec:results}
Figure \ref{fig1-Tprofiles} presents the magnitudes and phases of the temperature fields obtained by the Landauer framework [Eq. (\ref{deltaT_FIGMA})] and non-grey BTE [Eqs. (\ref{PBTE1})--(\ref{PBTE3})] conveniently plotted relative to the exact diffusive solution (\ref{PFourier}). BTE solutions display a gradual recovery towards Fourier diffusion with decreasing heating frequency, as physically appropriate. Landauer solutions, on the other hand, systematically maintain severe deviations at all frequencies below 100$\,$MHz. This seemingly puzzling behaviour arises from the fact that the Landauer solution does not properly converge to regular diffusive transport. Indeed, even when $\omega_{\text{H}}\tau_{\omega} \ll 1$ for most phonon modes, Eq. (\ref{deltaT_FIGMA}) still remains a weighted sum of exponentials with different decay rates. The Fourier solution (\ref{PFourier}), by contrast, is a single exponential but with decay rate and prefactor depending on the weighted sum $D \equiv \sum \frac{C_{\omega}}{C} \cdot D_{\omega}$. Phonon channels with limited diffusivity (such as those found in the optical branches) contribute sharply decaying exponentials with large prefactor to the Landauer solution, and thereby induce the surface temperatures to be overestimated by up to an order of magnitude.
\myfig[!htb]{width=0.45\textwidth}{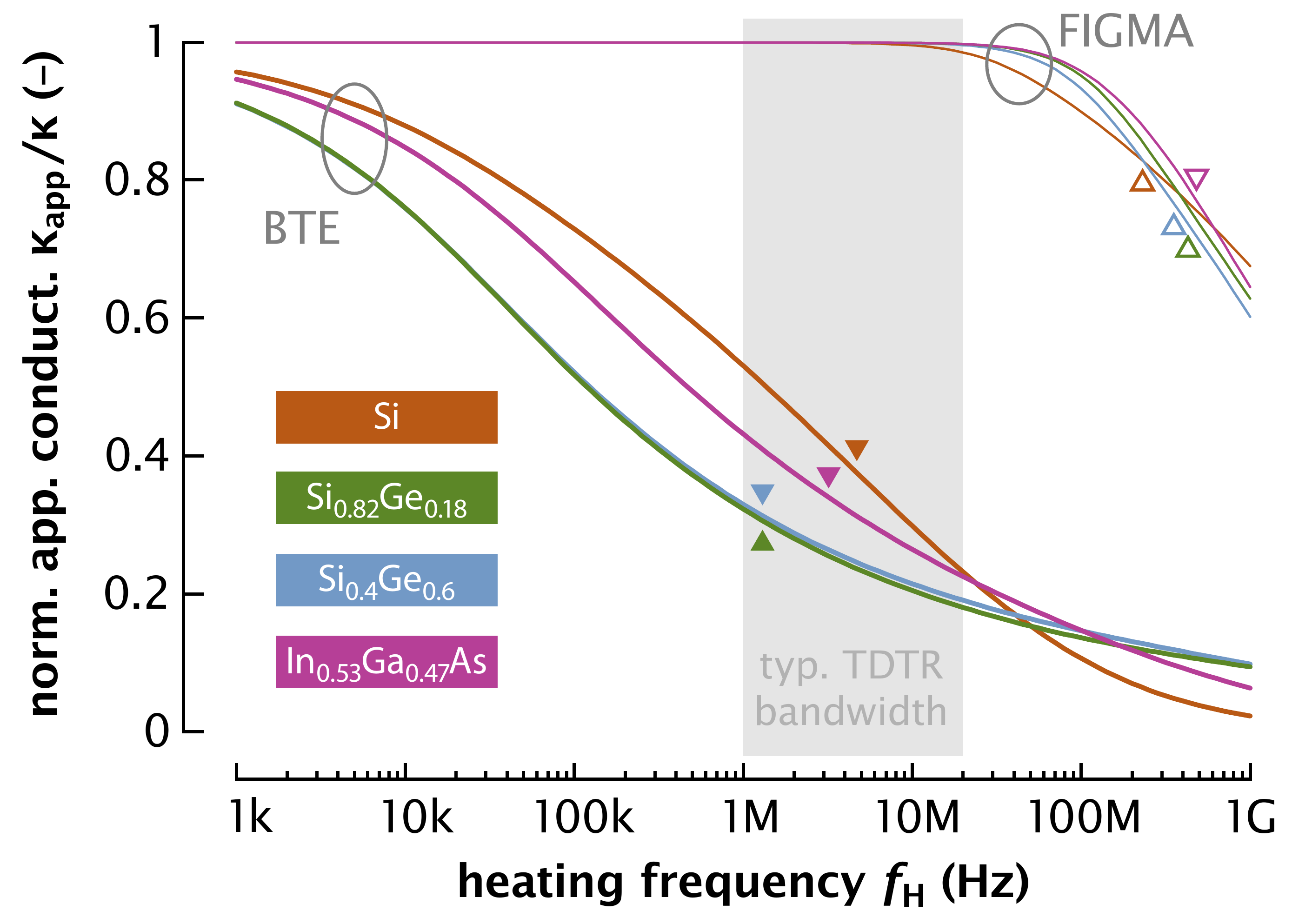}{Apparent conductivity suppression as observed at the semiconductor surface. $\blacktriangle$ and $\vartriangle$ mark characteristic frequencies for weakly and strongly quasiballistic effects respectively.}
\par
Figure \ref{fig2-kappa_surf} shows the apparent conductivities (\ref{kappaeffFIGMA}) and (\ref{kappaeffBTE}) normalised to the nominal Fourier value. We again observe substantial discrepancies between FIGMA and BTE solutions. Conductivity suppression in the nongrey BTE is induced by `lengthscale' effects while kinetic theory only captures a `timescale' effect, as conveyed by the marked characteristic heating frequencies
\begin{eqnarray}
& &\text{weakly quasiballistic effects:} \nonumber\\
& & \qquad \ell(\omega_{\text{H}}) / \Lambda_{\text{dom}} = 1 \quad \leftrightarrow \quad f_{\Lambda} = \frac{D}{\pi \Lambda_{\text{dom}}^2} \\
%\end{eqnarray}
%\begin{eqnarray}
& & \text{strongly quasiballistic effects:} \nonumber\\
& & \qquad \omega_{\text{H}} \, \tau_{\text{dom}} = 1 \quad \leftrightarrow \quad f_{\tau} = \frac{1}{2 \pi \tau_{\text{dom}}}
\end{eqnarray}
where we introduced `dominant' phonon metrics
\begin{equation}
\Lambda_{\text{dom}} = \sum \frac{\kappa_k}{\kappa} \, \Lambda_{k} \, |\cos \theta_k| \quad , \quad \tau_{\text{dom}} = \sum \frac{\kappa_k}{\kappa} \, \tau_k
\end{equation}
Our first-principles computations produce values on the order of $\tau_{\text{dom}} \simeq 0.5\,$ns and $\Lambda_{\text{dom}} \simeq 1\,\mu$m (precise values for each compound are listed in Appendix \ref{app:abinitiodata}). We attribute the far larger values ($\tau \simeq 130\,\text{ns} \leftrightarrow \Lambda \simeq 300\,\mu\text{m}$) quoted for Si$_{0.4}$Ge$_{0.6}$ by Ref. \onlinecite{ref1} partly to the simplified phonon dispersion and scattering law employed therein and partly to the $24 \times 24 \times 24$ wavevector grid we utilised. The `maximum' MFP and relaxation time is effectively capped by the grid resolution around the BZ center. However, we have verified that denser grids in fact worsen the discrepancy between FIGMA and BTE solutions: both $\Lambda_{\text{dom}}$ and $\tau_{\text{dom}}$ increase as expected but in slightly uneven proportion, causing the mismatch in onset frequencies $f_{\tau}/f_{\Lambda}$ to rise. We additionally remind that the main subject of scrutiny here is not the absolute accuracy of first-principles data, but rather the substantial discrepancy between FIGMA and BTE solutions for a common set of phonon properties.
\par
One is easily tempted to directly compare results from Fig. \ref{fig2-kappa_surf} to experimental values obtained by TDTR. However, three important aspects of the actual measurement are not yet captured by the investigated 1D configuration: (i) the experiment requires a metal transducer and therefore only probes the semiconductor indirectly; (ii) the experimental heat source has a Gaussian-shaped cross-section, bringing lateral heat spreading effects into play; and (iii) the experiment subjects the sample to modulated pulse trains rather than a pure sinusoid. Incorporation of these effects into 3D BTE treatments is well under way and may be the topic of a future publication. Our preliminary first-principles TDTR simulations of semiconductor alloys in the 1--20$\,$MHz range produce apparent conductivities up to twice as large as those observed in Fig. \ref{fig2-kappa_surf}, and in reasonable agreement with measurements. We stress that the mismatch between apparent conductivities inferred by TDTR and those observed at the semiconductor surface is not a computational error but rather constitutes an inherent artifact of the conventional `modified Fourier' interpretation of the raw TDTR data. Although a diffusive framework manages to fit the transient signals recorded at the transducer surface, it offers a poor representation of the quasiballistic semiconductor dynamics \cite{levy2}.
\par
We also see that the BTE predicts notable conductivity suppression in Si as well, contrary to experimental observations \cite{cahill,levy2}. This anomaly has been observed previously \cite{minnichVRMC} and may possibly be related to inherent limitations of the RTA \cite{collective} and/or interplay with the transducer \cite{wilson}. Resolving this open issue falls outside the scope of the present analysis but deserves further investigation.
\section{Discussion}\label{sec:discussion}
The detailed comparisons above reveal that FIGMA-based approaches fail to provide adequate approximations of temporally periodic BTE solutions. These findings stand in stark (and potentially surprising) contrast to previous reports \cite{collins,maassen_steadystate,maassen_transient} of good FIGMA performance in other quasiballistic transport settings.
\par
Maassen and Lundstrom demonstrated good agreement between Landauer and non-grey BTE solutions for both steady-state \cite{maassen_steadystate} and transient \cite{maassen_transient} temperature fields inside thin (3--300$\,$nm) Si films. However, heat conduction in nanoscaled slab structures is dominated by geometric constraints (boundary scattering), and therefore Refs. \onlinecite{maassen_steadystate} and \onlinecite{maassen_transient} do not offer representative evidence of the Landauer framework's suitability to describe quasiballistic transport in (semi)infinite media. 
\par
Collins and coworkers \cite{collins} demonstrated FIGMA to be a highly adequate approximation for 1D TTG. However, here too this outcome is not automatically portable to periodic heating configurations. To see why, it is worth reminding in this context that both TTG and TDTR probe the weakly quasiballistic regime. That is, in both configurations the Fourier diffusion paradigm breaks down because the characteristic length scale of the thermal gradient (grating period $\lambda_{\text{H}}$ and penetration length $\ell(\omega_{\text{H}})$ respectively) becomes comparable with phonon MFPs. Now, the grey-medium response to a \textit{spatially} periodic heat source naturally induces a conductivity suppression function that depends on the $\Lambda/\lambda_{\text{H}}$ ratio \cite{collins}. This thus captures the dominant quasiballistic effect and as a result, FIGMA extension is capable to provide an accurate description of conductivity suppressions observed in TTG experiments. In a display of mathematical and physical symmetry, the grey-medium response to a \textit{temporally} periodic heat source is characterised, as we saw above, by a conductivity suppression function that depends on $\tau/\tau_{\text{H}}$ (with $\tau_{\text{H}} = \omega_{\text{H}}^{-1}$ again the source period). The FIGMA model thus describes a GHz-range strongly quasiballistic effect but is left unable to capture the weakly quasiballistic conductivity suppressions at far lower frequencies that are inherently present in the BTE.
\par
Finally, it is interesting to note that lengthscale- and timescale-induced quasiballistic effects not just correspond to well separated threshold frequencies but in fact are connected to a profound distinction of how heat flux $q$ relates to the temperature gradient $Gr \equiv \partial \Delta T / \partial x$.
\par
Grey solutions are rigorously governed by the Cattaneo law $(1+s\tau) q = -\kappa \, Gr$ \cite{maassen_transient}, which we can recast as
\begin{eqnarray}
q(x,s) & = & -\kappa^{\ast}(s) \, Gr(x,s) \nonumber \\
\leftrightarrow \quad q(x,t) & = & - \int \limits_{0}^{t} \kappa^{\ast}(t') \, Gr(x,t-t') \, \mathrm{d}t'
\end{eqnarray}
The constitutive law has acquired temporal memory: the heat flux at a given time is codependent on the temperature gradient at earlier times through the convolution kernel $\kappa^{\ast}(t') = (\kappa/\tau) \exp(-t'/\tau)$. Non-grey BTE solutions (\ref{PBTE1}), by contrast, can be shown to correspond to
\begin{eqnarray}
q(\xi,t) & = & -\kappa^{\ast}(\xi) \, Gr(\xi,t) \nonumber \\
\leftrightarrow \quad q(x,t) & = & - \int \limits_{-\infty}^{\infty} \kappa^{\ast}(x') \, Gr(x-x',t) \, \mathrm{d}x'
\end{eqnarray}
The constitutive law has acquired `spatial memory', i.e. it has become delocalised: the heat flux at a given place is codependent on the temperature gradient at other locations through a convolution kernel $\kappa^{\ast}(x')$.
\section{Conclusions}\label{sec:conclusions}
In summary, we have analysed quasiballistic thermal transport in semi-infinite semiconductors under periodic heating regime with first-principles phonon dispersions and scattering rates. Comparing approximate models based on kinetic and Landauer theory to semi-analytic BTE solutions reveal that FIGMA-based approaches, in spite of excellent performance for spatially periodic heat sources, are ill suited to describe temporally periodic quasiballistic transport.
\section*{Acknowledgements}
BV acknowledges funding from the \textsc{alma} Horizon 2020 project (European Union Grant No. 645776) and thanks Natalio Mingo and Jes\'us Carrete (CEA-Grenoble) for helpful discussions and providing force constants.
\appendix
\section{Grey media revisited}\label{app:greymedia}
\subsection{Semi-infinite geometry}
The linearised RTA-BTE for the `forward' ($+$) and `backward' ($-$) propagating phonon modes in a grey medium with bulk heat capacity $C$ reads
\begin{equation}
\frac{\partial g^{\pm}}{\partial t} \pm |v_x| \frac{\partial g^{\pm}}{\partial x} = -\frac{g^{\pm} - \frac{1}{2} \, C \, \Delta T}{\tau}
\end{equation}
We now consider a semi-infinite geometry extending over $x \geq 0$ in equilibrium at $t=0$ and carry out Laplace transformations with respect to both space ($x \leftrightarrow \sigma$) and time ($t \leftrightarrow s$). Accounting for the fact that $\partial f(x)/\partial x \leftrightarrow \sigma F(\sigma) - f(x=0)$ and introducing $\Lambda_x = |v_x| \tau$ we find
\begin{equation}
g^{\pm}(\sigma,s) = \frac{\frac{1}{2} \, C \, \Delta T(\sigma,s) \pm \Lambda_x \, g^{\pm}(x=0,s)}{1 + s \tau \pm \sigma \Lambda_x} \label{greygplusminus}
\end{equation}
We subject the top surface to a source heat flux: $|v_x| \cdot [g^+(x=0,s) - g^-(x=0,s)] = q_S(s)$. From (\ref{greygplusminus}) we can now derive the total deviational energy density $P = g^+ + g^- = C \, \Delta T$:
\begin{equation}
P(\sigma,s) = \frac{\tau \, (1+s\tau) \, q_S(s) - \sigma \, \Lambda_x^2 \, P_0(s)}{s\tau \, (1+s\tau) - \sigma^2 \Lambda_x^2}
\end{equation}
where $P_0(s) \equiv g^+(x=0,s) + g^-(x=0,s)$. Notice that $\sigma \, P(\sigma,s) \rightarrow P_0(s)$ for $\sigma \rightarrow \infty$, as appropriate. The net heat flux $q = |v_x| \, (g^+-g^-)$ inside the medium immediately follows from energy conservation considerations:
\begin{eqnarray}
& & \frac{\partial q}{\partial x} + \frac{\partial P}{\partial t} = 0 \quad \leftrightarrow \quad \sigma \, q(\sigma,s) - q_S(s) + s\, P(\sigma,s) = 0\nonumber \\
& & \quad \Rightarrow \,\, q(\sigma,s) = \frac{q_S - s P}{\sigma} = \frac{\Lambda_x^2 \, [s \, P_0(s) - \sigma \, q_S(s)]}{s\tau \, (1+s\tau) - \sigma^2 \, \Lambda_x^2}
\end{eqnarray}
Noting that the temperature gradient $Gr(x,t) \equiv \partial \Delta T(x,t)/\partial x$ reads $Gr(\sigma,s) = [\sigma\,P(\sigma,s) - P_0(s)]/C$ in transformed domains, the apparent conductivity $\kappa_{\text{app}}(\sigma,s) \equiv \|-q(\sigma,s)/Gr(\sigma,s) \|$ is found to be
\begin{equation}
\kappa_{\text{app}}(\sigma,s) = \left\| \frac{-C \, q(\sigma,s)}{\sigma \, P(\sigma,s) - P_0(s)} \right\| =  \left\| \frac{C \, \Lambda_x^2}{\tau \, (1+s\tau)} \right\|
\end{equation}
The vanishing of $\sigma$ indicates that heat flux and temperature gradient have identical spatial signatures, as observed in Ref. \onlinecite{ref1}. Under periodic heating $s = j \omega_{\text{H}}$, the semi-infinite grey medium (having nominal conductivity $\kappa \equiv C \, |v_x| \, \Lambda_x = C\, \Lambda_x^2 / \tau$) is thus characterised by
\begin{equation}
\kappa_{\text{app}}(\omega_{\text{H}}) = \left\| \frac{\kappa}{1 + j \omega_{\text{H}} \tau}\right\| = \frac{\kappa}{\sqrt{1 + (\omega_{\text{H}} \tau)^2}} \label{kappagreyappendix}
\end{equation}
as mentioned in the main text.
\par
It is worth pointing out that the thermal field inside the semi-infinite grey medium is \textit{exactly twice} that of the infinite counterpart, even in (quasi)ballistic regimes. To prove this, consider the symmetrically extended energy density $\hat{P}(x,t) = P(|x|,t)$ where $x$ spans the entire real axis. In Fourier-Laplace domain ($x \leftrightarrow \xi , s \leftrightarrow t)$, the Green's function $\hat{G}(\xi,s) \equiv \hat{P}(\xi,s)/q_S(s)$ is readily connected to the previously derived semi-infinite solution:
\begin{eqnarray}
\hat{G}(\xi,s) & = & \frac{P(\sigma = j\xi,s) + P(\sigma = -j\xi,s)}{q_S(s)} \nonumber \\
& = & \frac{2 \tau \, (1+s\tau)}{s\tau \, (1+s\tau) + \xi^2 \Lambda_x^2}
\end{eqnarray}
This precisely equals two times the single pulse response (\ref{Pxisgrey}) of the infinite grey medium (derived below).
\myfigwide[!htb]{width=0.84\textwidth}{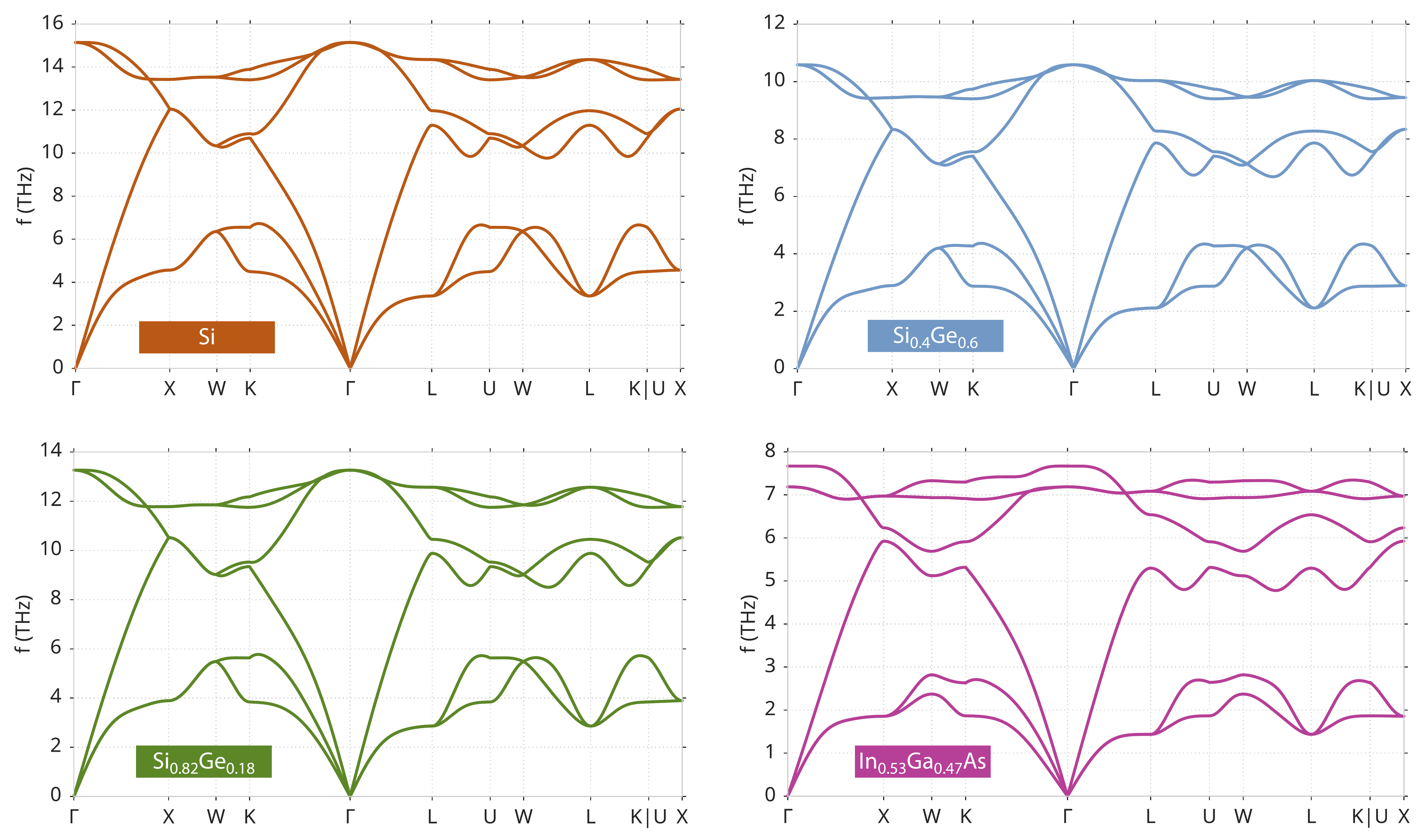}{Phonon dispersion curves obtained from first-principles atomic force constant computations.}%
\subsection{Infinite geometry}
The single pulse energy density response $P(\xi,s)$ in Fourier-Laplace domain is immediately found by evaluating the general (non-grey) solution of the BTE (provided by Ref. \onlinecite{levy1}) for a single phonon pair:
\begin{eqnarray}
& & P(\xi,s) = \frac{\tau \, \Xi(\xi,s)}{1-\Xi(\xi,s)} \quad \text{with} \quad \Xi = \frac{1+s\tau}{(1+s\tau)^2 + \xi^2 \Lambda_x^2} \nonumber \\
\Rightarrow \,\, & & P(\xi,s) = \frac{\tau (1+s \tau)}{s \tau (1+s \tau) + \xi^2 \Lambda_x^2} \label{Pxisgrey}
\end{eqnarray}
One can also derive this solution explicitly by using Maassen and Lundstrom's observation \cite{maassen_transient} that thermal transport in a grey medium rigorously obeys the hyperbolic heat equation at all length and time scales:
\begin{equation}
\frac{\partial P(x,t)}{\partial t} + \tau \, \frac{\partial^2 P(x,t)}{\partial t^2} - D \, \frac{\partial^2 P(x,t)}{\partial x^2} = 0
\end{equation}
where $D = \Lambda_x^2/\tau$ denotes the diffusivity. The single pulse response corresponds to initial condition $P(x,t=0) = \delta(x) \leftrightarrow P(\xi,t=0) = 1$ so that transforming the hyperbolic heat equation produces
\begin{equation}
[s P(\xi,s) - 1] + \tau \, [s^2 P(\xi,s) - s \cdot 1] + \frac{\Lambda_x^2}{\tau} \, \xi^2 \, P(\xi,s) = 0
\end{equation}
This again yields the solution (\ref{Pxisgrey}).
\section{Ab-initio phonon properties}\label{app:abinitiodata}%
First-principles phonon dispersions, bulk thermal properties and normalised cumulative conductivity curves (`MFP spectra') $\kappa_{\Sigma}(\Lambda)/\kappa$ are provided in Fig. \ref{fig3-dispersions}, Table \ref{tab:abinitioprop} and Fig. \ref{fig4-kappacumul} respectively.
\begin{table}[!htb]
\caption{Thermal properties at 300$\,$K obtained from first-principles phonon dispersions and scattering rates.}\label{tab:abinitioprop}
\vspace{4mm}%
\begin{tabular}{ccccc}
\hline
Compound & $\kappa$ & $C$ & $\Lambda_{\text{dom}}$ & $\tau_{\text{dom}}$ \\
 & [W/m-K] & [MJ/m$^3$-K] & [$\mu$m] & [ns] \\
\hline
Si & 156 & 1.626 & 2.59 & 0.69\\
Si$_{0.4}$Ge$_{0.6}$ & 7.12 & 1.654 & 1.04 & 0.45\\
Si$_{0.82}$Ge$_{0.18}$ & 7.56 & 1.660 & 1.04 & 0.37\\
In$_{0.53}$Ga$_{0.47}$As & 8.28 & 1.568 & 0.72 & 0.33\\
\hline
\end{tabular}%
\end{table}%
\section{Variance-reduced Monte Carlo simulations of semi-infinite media}\label{app:VRMC}
Being a time-stepping technique, VRMC cannot simulate periodic heating regimes directly. However, it is ideally suited to obtain the time-domain single pulse response, which entirely and unambiguously characterises the transient transport dynamics of the system. As heat is injected at the top surface only at $t=0$ and convective/radiative cooling is ignored, the top surface effectively acts as an adiabatic boundary throughout the rest of the simulation: $q_{\text{net}}(x=0,t>0) = 0$. Two possiblities (with relative occurrences governed by the surface specularity $p$) now arise for how deviational particles interact with this adiabatic wall (Fig. \ref{fig5-boundaryscattering}). If the particle reflects specularly, the wall simply acts as a perfect mirror for the $x$-coordinate of the trajectory with total travel time and distance left unchanged. If all particles behave this way ($p \rightarrow 1$), the solution for the semi-infinite medium is exactly twice that for the infinite structure.
\par
With realistic semiconductor surfaces neither perfecly smooth nor perfectly rough, we have used $p=0.5$ for our simulations. Each particle interacting with the wall thus has a 50\% chance of being scattered diffusely. In such cases, we terminate the trajectory of the incoming particle at the time and location of impact, and randomly draw a new phonon mode from those whose $x$-projected velocity has the same magnitude as the incoming mode but opposite sign: $v_x^{\text{in}} + v_x^{\text{out}} = 0$. This ensures that the `forward' and `backward' energy fluxes at the top surface are always in balance, precisely what is mandated by an adiabatic boundary. In a discretised wavevector grid, velocity magnitudes are unlikely to be precisely equal within machine precision except for modes that are equivalent through crystal symmetries, which would simply lead to specular behaviour again. We therfore allow a generous $\pm$1\% tolerance for the criterion, i.e. we randomly select from modes that satisfy $|v_x^{\text{in}} + v_x^{\text{out}}| \leq 0.01\,|v_x^{\text{in}}|$. As sketched in Fig. \ref{fig5-boundaryscattering}, outgoing modes can differ substantially from the incoming one in terms of both scattering rate and angle with the surface normal. Thus, their contributions to the 1D thermal field can be quite different from their specular counterpart.
\par
Despite the presence of diffuse boundary scattering effects, the simulated temperature field can still be excellently approximated by simply doubling the solution for the infinite structure, as revealed by Fig. \ref{fig6-VRMC}.
\myfig[!htb]{width=0.45\textwidth}{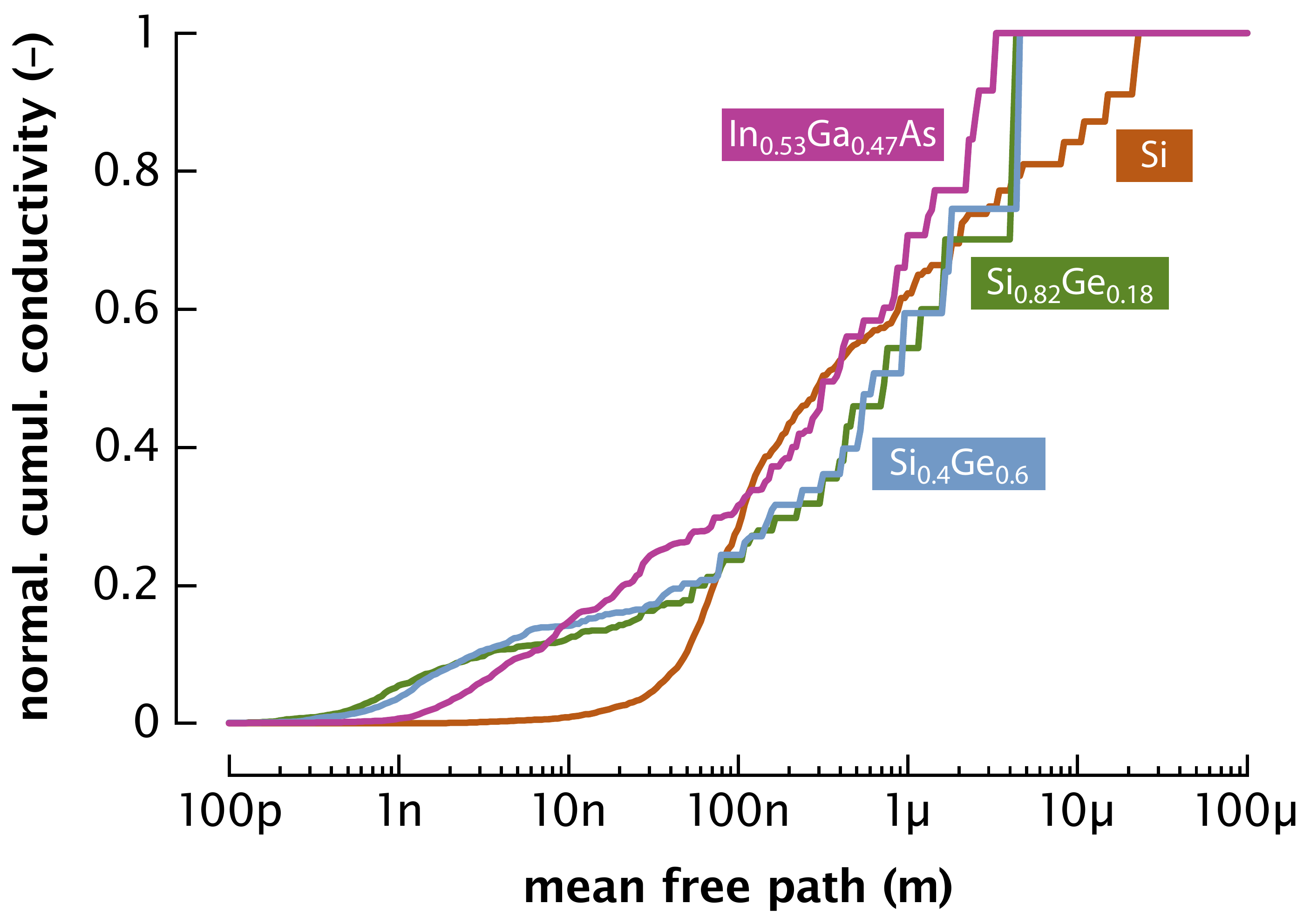}{Normalised cumulative conductivity curves obtained from first-principles phonon dispersions and scattering rates.}
\pagebreak[4]
\myfig[!htb]{width=0.4\textwidth}{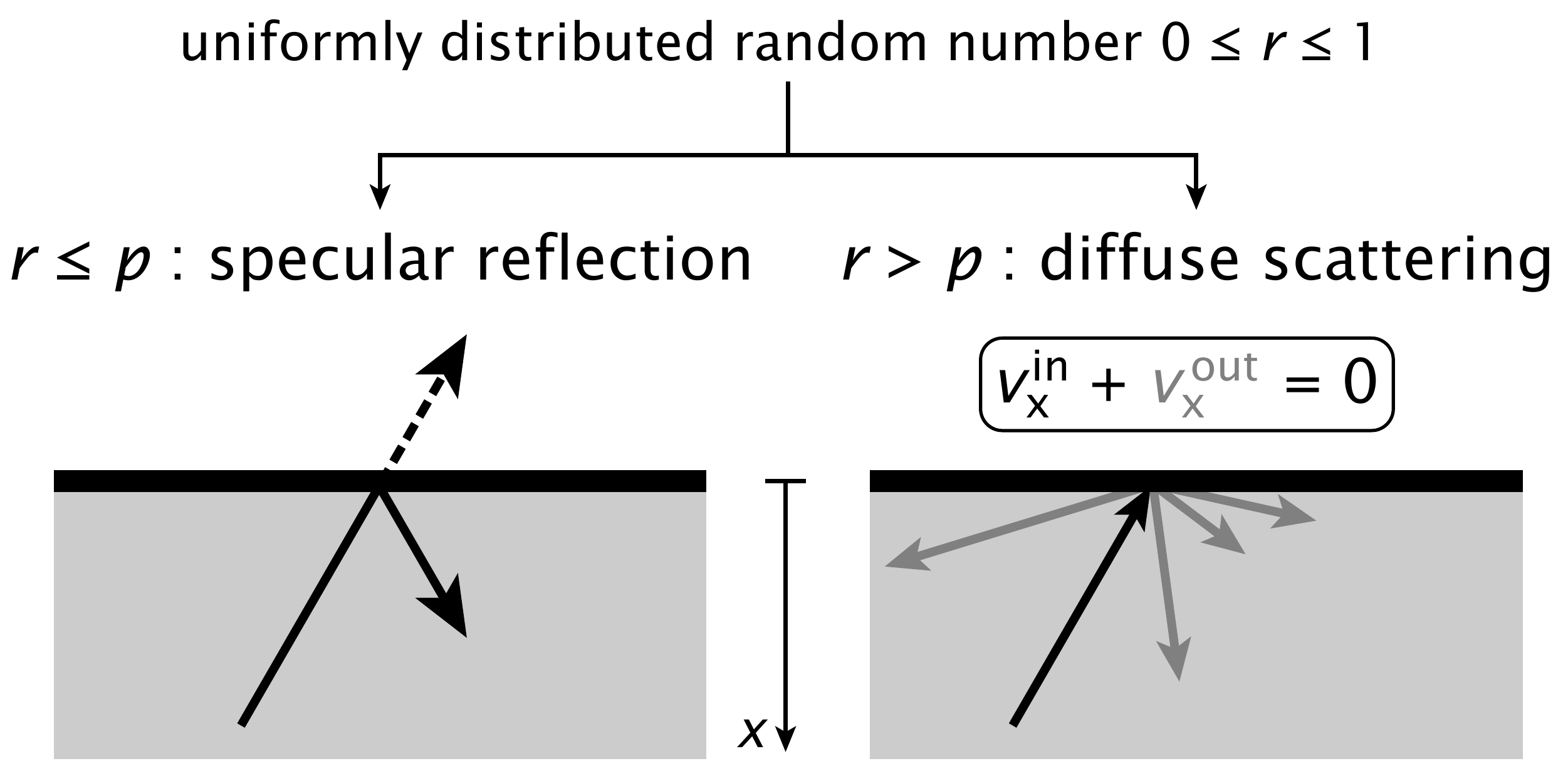}{Interaction of Monte Carlo particles with the top surface (having specularity $p$) in a semi-infinite medium.}
\myfigwide[!htb]{width=\textwidth}{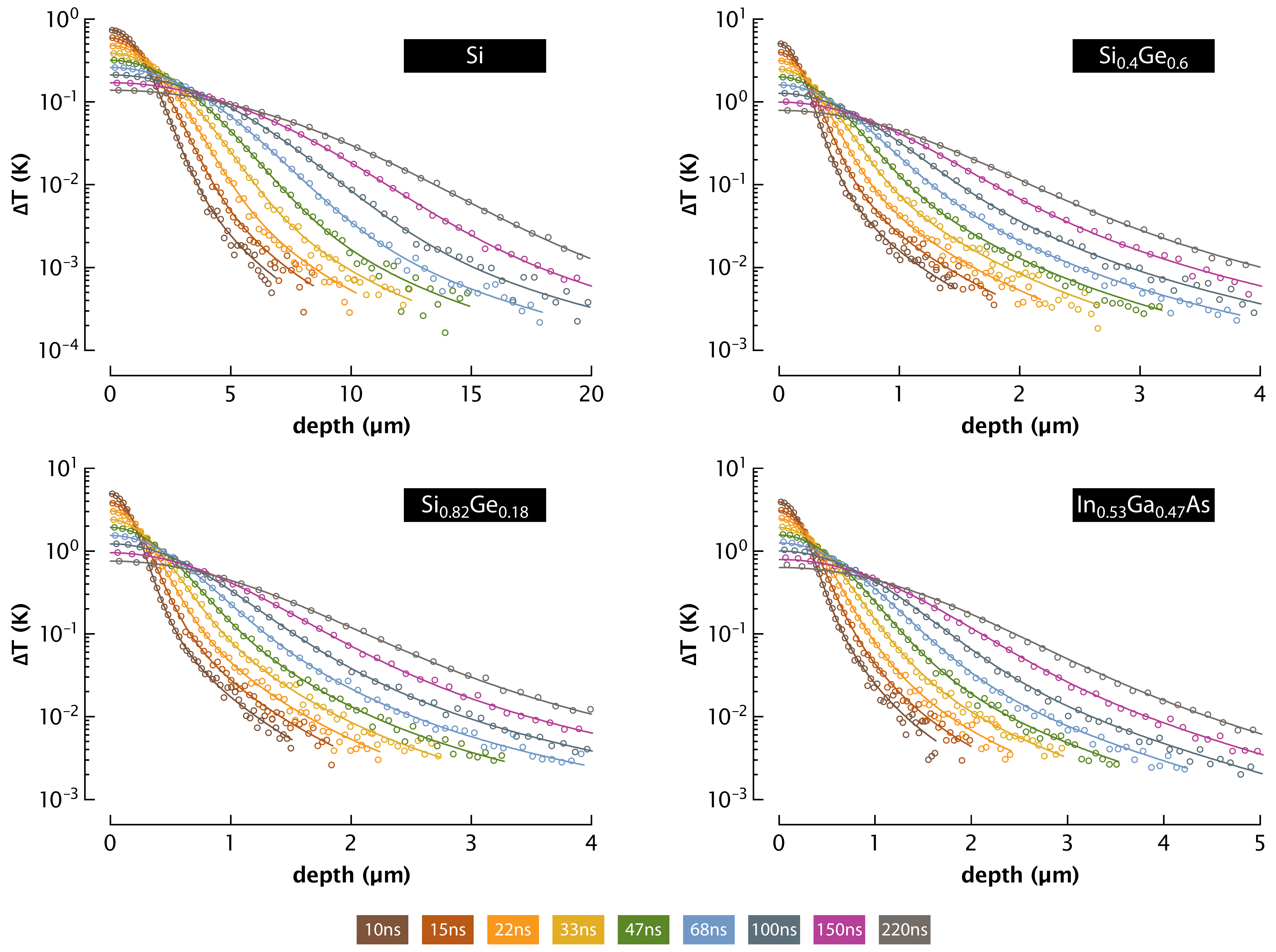}{Single pulse response in semi-infinite media with first-principles phonon dispersions and scattering rates. Symbols: Variance-reduced Monte Carlo simulations with $10^5$ deviational particles for surface specularity $p=0.5$. Lines: semi-analytic BTE solution $\Delta T(x,t) = (2/C)\,P(x,t)$ calculated with the procedures described in Section \ref{sec:BTE}.}%
%\bibliography{biblio}
% INSERTED BBL FILE
%merlin.mbs apsrev4-1.bst 2010-07-25 4.21a (PWD, AO, DPC) hacked
%Control: key (0)
%Control: author (8) initials jnrlst
%Control: editor formatted (1) identically to author
%Control: production of article title (-1) disabled
%Control: page (0) single
%Control: year (1) truncated
%Control: production of eprint (0) enabled
%
\end{document}